\documentclass[aps,pra,twocolumn,superscriptaddress,showpacs,floatfix]{revtex4-1}
\usepackage{graphicx,amsmath,amssymb,xspace,epsfig,float}

\usepackage{color} 
\usepackage{latexsym}
\usepackage{bm}
\usepackage{wasysym}

\usepackage{graphicx}
\usepackage{epstopdf}
\graphicspath{{./}{./Figs/}{./Figs_paper/}}

\usepackage[colorlinks,linkcolor=blue,urlcolor=blue,bookmarks=true,pdftitle={}]{hyperref}

\newcommand{\mass}{\mathsf{m}}

\newcommand{\eexp}[1]{\mathrm{e}^{#1}}

\newcommand{\braket}[1]{\left\langle #1 \right\rangle }

\newcommand{\be}[1]{\begin{eqnarray} \label{e#1}}
\newcommand{\beq}{\begin{eqnarray}}
\newcommand{\eeq}{\end{eqnarray}} 

\newcommand{\hide}[1]{}

\newcommand{\Eq}[1]  {{\textcolor{blue}{Eq.}}~(\ref{#1})} 
\newcommand{\Fig}[1] {{\textcolor{blue}{Fig.}}~\ref{#1}}

\newcommand{\Cn}[1]{\begin{center} #1 \end{center}}
\newcommand{\hrefl}[1]{\href{#1}{[link]}}

\makeatletter
\renewcommand*{\@fnsymbol}[1]{\ifcase#1\or \textdagger \else\@arabic{#1}\fi}
\makeatother

\begin{document}

\title{Resonant persistent currents for ultracold bosons on a lattice ring}

\author{Geva Arwas}
\affiliation{Department of Physics, Ben-Gurion University of the Negev, Beer-Sheva 84105, Israel}

\author{Doron Cohen}
\affiliation{Department of Physics, Ben-Gurion University of the Negev, Beer-Sheva 84105, Israel}

\author{Frank Hekking} \thanks{Deceased on May 15, 2017.}
\affiliation{Univ. Grenoble Alpes, CNRS, LPMMC, 38000 Grenoble, France}

\author{Anna Minguzzi}
\affiliation{Univ. Grenoble-Alpes, CNRS, LPMMC, 38000 Grenoble, France}


\begin{abstract}
We consider a one-dimensional bosonic gas on a ring lattice, in the presence of a localized barrier, and under the effect of an artificial gauge field. By means of exact diagonalization we study the persistent currents at varying interactions and barrier strength, for various values of lattice filling. While generically the persistent currents are strongly suppressed in the Mott insulator phase,  they show a resonant behaviour when the barrier strength becomes of the order of the interaction energy. We explain this phenomenon using an effective single-particle model. We show that this effect is robust at finite temperature, up the temperature scale where persistent currents vanish.
\end{abstract}

\maketitle



\section{Introduction}

Ultracold atoms on ring traps are the object of an active experimental investigation, and provide a novel generation of quantum simulators. The ring geometry avoids the inhomogeneities of the harmonic confinement, as well as the effects of sharp boundary conditions, typical e.g. of box confinements.  A very natural observable in the ring geometry is the circulating current, which is experimentally accessible with ultracold atoms by time-of-flight  \cite{first-nist-experiments,hadzibabic} or  by spiral interferogram techniques \cite{dalibard,nist}.

If the radial  confinement along the ring is very tight, i.e., when all the energy scales in the problem are smaller than the radial confinement energy, the ring can be effectively described as one-dimensional. 
The exact wavefunction  of $N$~interacting bosons in a one-dimensional ring of length~$L$  with contact interactions has  
been found by Lieb and Liniger \cite{Lieb-Liniger}. The solution is controlled by the dimensionless parameter 
\be{100}   
\gamma \ \ = \ \ \frac{g \mass L}{\hbar^2 N}
\eeq
where $\mass$ is the mass of the particles, and the interaction is ${u_{int}(x-x')=g\delta(x-x')}$. 
For ${\gamma\ll 1}$ the system can be described to a good approximation by the Gross-Pitaevskii equation,
which we regard as the {\em classical} limit, 
while for ${\gamma\gg 1}$ the bosons reach the impenetrable limit and undergo  fermionization \cite{Girardeau}, i.e. share the same nodes as the wavefunction of a noninteracting Fermi gas in the same external potential.
Irrespective of the interaction strength $\gamma$, for ultracold bosons on a ring it has been argued that 
in the presence of a gauge field $\Phi$, the persistent 
current $I(\Phi)$ exhibits full Aharonov-Bohm oscillations \cite{Leggett-theorem,mueller-groeling,loss}. 
This is no longer true if a barrier is introduced.

A device that incorporates a barrier (or a weak link), 
and a gauge field $\Phi$, i.e. the analog of the rf-SQUID in superconductors, can be realized by rotating a laser-induced 
optical potential. Another way to induce a gauge field is via laser-assisted transitions. 
Thanks to the barrier, it is possible to induce persistent currents 
of controllable amplitude \cite{opt-current}. Such setup can be also used to realize 
a qubit, based on superposition of current states, 
analogous of a flux qubit in superconductors \cite{moij,agamalyan-cominotti,muntsa1,muntsa2}.
It has been established \cite{opt-current} that in such device the persistent 
current amplitude has a non-monotonic dependence on $\gamma$.
In the Gross-Pitaevskii regime it increases due to a classical screening effect,
while in the large $\gamma$ regime it decreases due 
to quantum fluctuations. Thus, there is an optimal value 
of $\gamma$ for which the persistent current amplitude has a maximum.

There are several motivations to introduce an optical potential $V_L(x) = V_0 \cos^2(\pi x/a)$
to the ring. Naturally, it allows further control over the device characteristics. But furthermore, it has been suggested that such lattice geometry is essential for the purpose of coherent operations \cite{sfr}. 
Below we shall assume that the optical lattice can be effectively
described by the Bose-Hubbard Hamiltonian (BHH), 
which describes the motion of $N$ bosons along an $M$-site ring, where  $M=L/a$ (see \cite{sfa} and references within).
In the absence of interactions the bosons condense into the lowest wavelength orbital,
and the qualitative behaviour of the lattice ring is similar to the uniform-ring case,
with some effective mass  $\mass^*$ that is determined by the inter-site hopping frequencies.
Associated with the effective mass one can define an effective Lieb-Liniger parameter ${\gamma^*}$.
In the quantum domain (${\gamma^*>1}$) the system enters the strongly correlated regime,
and the ratio $N/M$ becomes important.  
In this regime the ring may undergo a transition towards a mesoscopic Mott insulator state.
For commensurate filling one may expect that the persistent current amplitude
would exhibit a non-monotonic dependence on $\gamma^*$,  
with a drastic drop within the Mott regime.

In the present work we  show that the transition between the Gross-Pitaevskii regime and the
Mott regime is in fact mediated by resonances where the persistent current amplitude
exhibits pronounced maxima.
The model system is introduced in Section~II,  
while the major numerical observations are displayed in Section~III, 
and explained in Section IV.  
The main features are captured by the single particle picture 
of Section~V.  The effect of temperature is examined in Section~VI.
We also discuss in Section~VII the related setup which incorporates 
a weak link, rather than a barrier.

\section{The model}

We consider a system of $N$ bosons confined to an $M$-site ring lattice. The system  is described  by the Bose Hubbard(BH) model:
\begin{eqnarray} 
\mathcal{H} &=& - J  \sum_{j=1}^{M} \left(\eexp{i(\Phi/M)} {a}_{j{+}1}^{\dag} {a}_{j} + \text{h.c.} \right) \nonumber \\
&+& W {a}_{1}^{\dag} {a}_{1} 
\ + \ \frac{U}{2}  \sum_{j=1}^{M} {a}_{j}^{\dag} {a}_{j}^{\dag} {a}_{j} {a}_{j} 
\label{e1}
\end{eqnarray}
where $J$ is the hopping energy, $U$ is the on-site interaction, 
and $j$ mod$(M)$ labels the sites of the ring. 
  ${a}_{j}$ and ${a}_{j}^{\dag}$ are the bosonic creation and annihilation operators, obeying the usual commutation relations $[a_j,a^\dag_\ell]=\delta_{j,\ell}$.
The phase $\Phi$ corresponds to the artificial gauge field applied to the system. In the case of a rotating barrier, it is proportional to 
the rotation frequency. In addition, a repulsive barrier of strength $W>0$ is located on the lattice site $j=1$. The total number of particles $N=\sum_j \langle  {a}_{j}^{\dag} {a}_{j} \rangle = \sum_j \langle {n}_j\rangle  $ is kept fixed in the calculation.
The persistent current is defined according to the thermodynamic identity
\begin{equation}
\label{e2}
\mathcal{I} \ \ = \ \  -\frac{\partial \braket{\mathcal{H}}}{\partial \Phi}
\end{equation}
In order to characterize the regimes of interactions and barrier strength, 
it is useful to introduce the following dimensionless parameters
\beq
u \ \ &=& \ \ M\frac{NU}{2J} \\ 
w \ \ &=& \ \ M\frac{W}{2J}
\eeq
These parameters naturally arise \cite{sfr} while considering the classical 
limit of the Bose-Hubbard model, where the dynamics is given by the discrete non-linear Schrodinger equation (DNLS).
The DNLS, which provides a classical description of the system, 
is in fact the discrete version of the Gross-Pitaevskii equation. 
The classical parameters $u$ and $w$ are obtained by rescaling the DNLS 
into a dimensionless form.   
The second-quantization of the DNLS introduces the effective Planck-constant $1/N$
that expresses the occupation-phase uncertainty.   

To gain further insight into the significance of~$u$ and~$w$ 
it is useful to link them to the dimensionless parameters 
of the continuous ring as defined in \cite{opt-current},  
where the two-body interaction potential is ${u_{int}(x-x')=g\delta(x-x')}$, 
and the localized barrier potential is taken as ${v_{barr}(x)=v\delta(x)}$.
In this limit, the natural dimensionless parameters are the Lieb-Liniger coupling 
strength $\gamma^*$, which is calculated using \Eq{e100} with the effective mass $\mass^*$, 
and the dimensionless barrier strength ${\lambda^*=v \mass^* L/(\pi\hbar^2)}$.
One readily deduces that $J=\hbar^2/(2\mass^* a^2)$, 
and $U=g/a$, and $W=v/a$. Accordingly 
\beq
\gamma^* \ \ &=& \ \ \left(\frac{1}{N}\right)^2  u \\
\lambda^* \ \ &=& \ \ \left(\frac{1}{\pi}\right) \ w
\eeq
For $\gamma^* > 1$ quantum fluctuations become important and the DNLS description fails to describe the system. At integer lattice fillings,
the Mott transition takes place in the thermodynamic limit. 
On a finite ring, a crossover towards a gapped, incompressible state takes place; as the number of lattice sites $M$ becomes larger the crossover becomes sharper. 
In this  work we are not taking the limit $M\to\infty$ 
since we want to explore the operation of a finite size circuit, and its persistent currents.

\section{The persistent current regime diagram}

\begin{figure*}
\begin{center}
\includegraphics{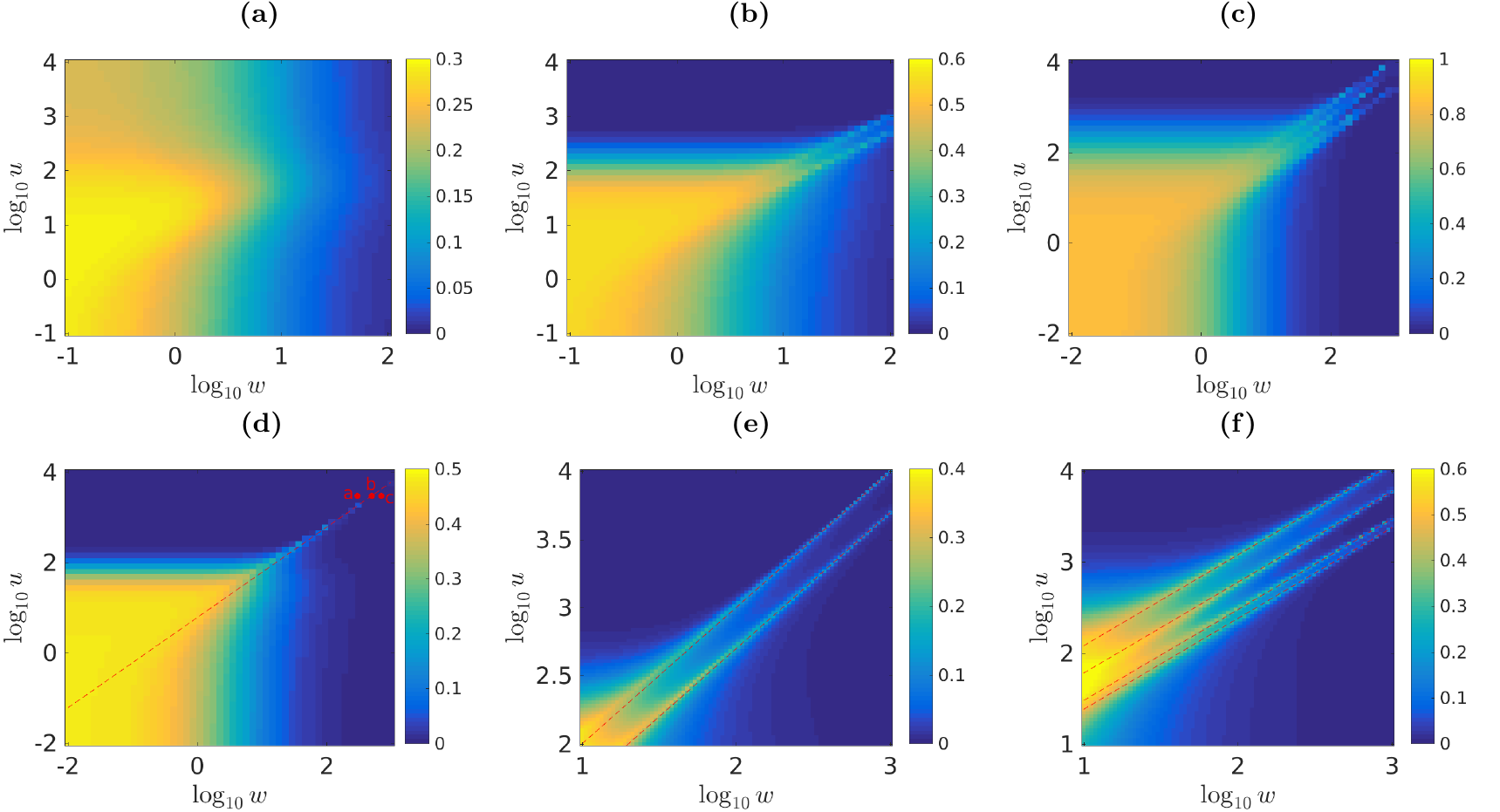}
\caption{ \label{fg1} The persistent current amplitude $\alpha$ as a function of $(w,u)$. Panels (a)-(d) are for rings with $(M,N)$ as follows: $(10,4)$ ; $(5,10)$ ; $(3,12)$ ; $(6,6)$. Panels (e) and (f) are zoomed versions of (b) and (c) respectively.
The value plotted is calculated using \Eq{alpha} where the maximal current $I(\Phi)$ is picked from all $\Phi \in [0,\pi] $ with step size of $0.05 \pi$. 
The dashed red lines are given by \Eq{e5}. The markers in (d) corresponds to the spectra in \Fig{fg2}. 
} 
\end{center}
\end{figure*}

We present now the results obtained from exact diagonalization for the persistent currents as a function of barrier and interaction strength. For each values of $(w,u)$ we calculate $I(\Phi)$ for the ground state and define the persistent current amplitude $\alpha$ via 
\beq \label{alpha}
\max\left[I(\Phi)\right] \ \ = \ \ 2J \frac{N}{M} \ \alpha(w,u) 
\eeq
such that full oscillations correspond to $\alpha=1$.  
The results for $\alpha(w,u)$ for several choices of $M$ and $N$ are illustrated in \Fig{fg1}. 

For filling smaller than one (panel a) the persistent current displays a non-monotonous behaviour as a function of interaction strength, thereby recovering the results of \cite{opt-current}. The position of the maximum increases with barrier strength.
For very large $w$ the ring is effectively disconnected by the barrier, and $\alpha$ vanishes. 
Otherwise the  current is non vanishing, even in the $u\to\infty$ limit: the ring does not become a Mott insulator.  

At commensurate fillings (panels b-d), the behavior is considerably different. 
The persistent current amplitude  $\alpha$ is strongly suppressed for large $u$ even for a small barrier,
indicating the onset of the mesoscopic Mott insulator state. 
But there is a twist on top of this observation:  for $U\sim W$ the persistent current
amplitude exhibits pronounced maxima. A zoom on this region shows that it consists of stripes 
whose number corresponds to the integer part of the filling $N/M$.

\section{The resonant structure of the persistent current amplitude}

We now provide an explanation for the resonant behaviour of the persistent-current amplitude.
Consider a system with $N$ bosons. For large interaction strengths, i.e. $U/J\gg 1$ it makes sense to 
represent the total number of particles as follows:
\beq \label{e9}
N \ = \ Mq_c + p \ = \ k + (M{-}1)q_k + p_k   
\eeq  
The occupation floor in the absence of a barrier is defined as the integer part of $N/M$, 
namely $q_c=\lfloor N/M \rfloor$. The reminder $p$ denotes that number of excess particles 
in the conduction band above the floor. The second equality defines the occupation floor $q_k$ and the excess particles $p_k$, 
given that $k$ particles reside at the barrier. Namely, 
\beq
q_k \ = \ \left\lfloor \frac{N-k}{M-1}  \right\rfloor
\ = \ q_c + \left\lfloor \frac{p+(q_c-k))}{M-1} \right\rfloor
\eeq
Note that $k=0,1,2,...$ is a non-negative 
integer, whose maximal value is~$N$. 
As the barrier is lowered to zero, $k$ increases from $0$ to $q_c$, while $q_k$ decreases 
from $q_0$ to $q_c$.

\begin{figure*}
\begin{center}
\includegraphics[width=0.94\hsize]{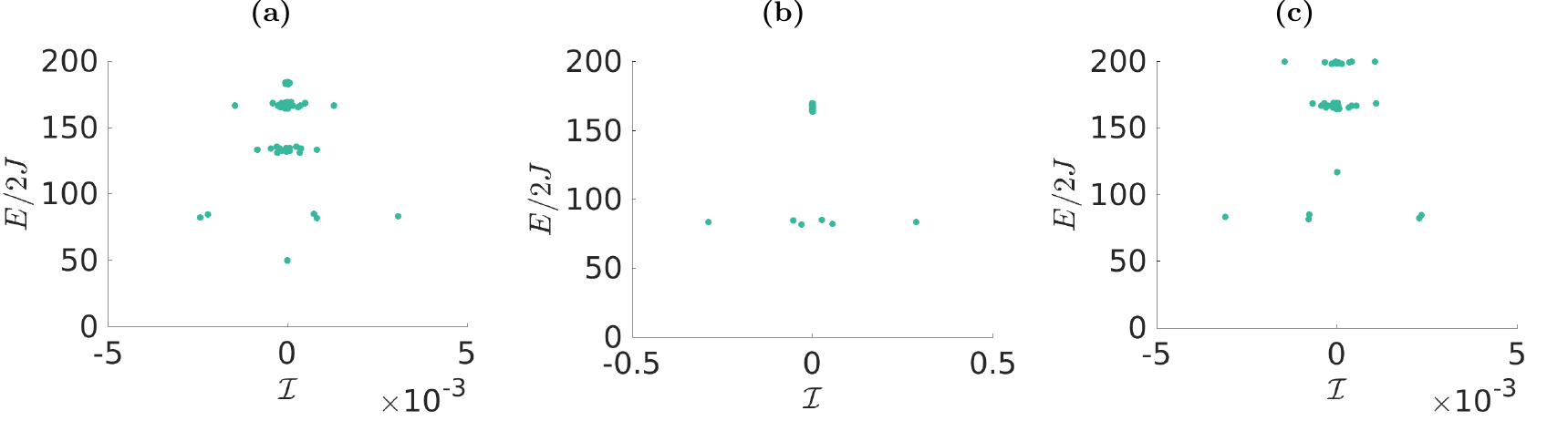}
\caption{ \label{fg2} Low energy spectrum for an $M=6$ ring with $N=6$ particles and $\Phi=0.9 \pi $. Each point represents an eigenstate, positioned according to its energy $E$ (vertical axis) and its current $\mathcal{I}$
(horizontal axis). The current is in units of $2NJ/M$, note the different axis limits in the middle panel. 
In all panels $u=3000$, while $w=300,500,700$ from left to right. This corresponds to the large $u$ regime, with $w$ values that are on the left,at the center, and to the right of the resonance. The $w$ values are marked in \Fig{fg1}(d). Note that we only show here states with $E<200$ and not the entire spectrum. 
} 
\end{center}
\end{figure*}

\begin{figure*}

\begin{center}
\includegraphics{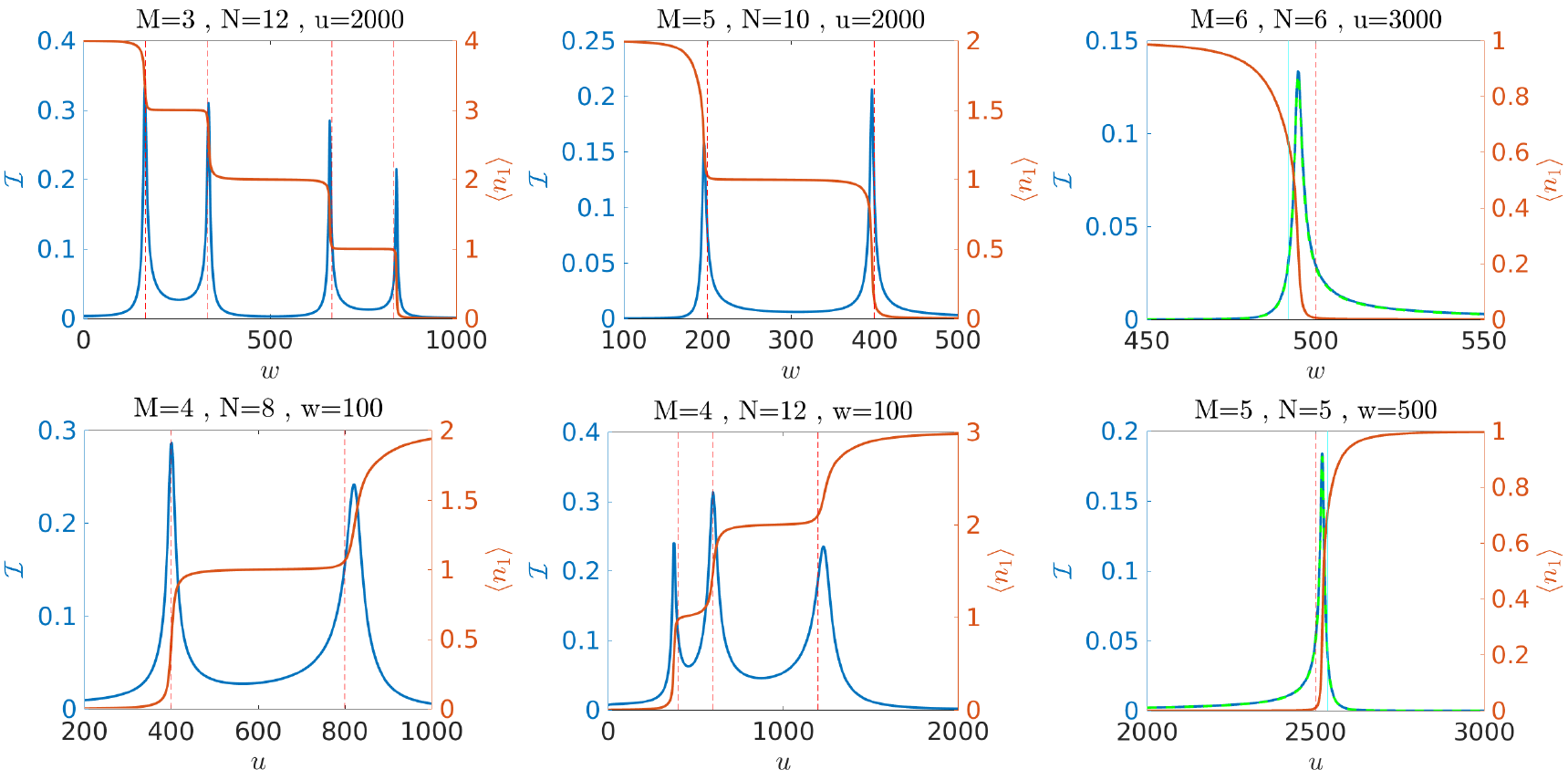}
\caption{ \label{fg2a} 
The ground-state current (blue spiky lines), in units of $2NJ/M$, and the barrier occupation (red staircase lines) as a function of dimensionless barrier strength $w$ (upper panels) and dimensionless interaction strength $u$ (lower panels). The vertical red lines indicate the expected location of the resonances, using \Eq{e5} (neglecting the $2J_q-\Delta$ shift). In the rightmost panels we also plot the current using the effective single particle Hamiltonian \Eq{e6} (green dashed), and the expected location of the resonances, including the $2J_q-\Delta$ correction (vertical cyan lines).
In all panels the flux is $\Phi=0.9 \pi$. 
} 
\end{center}
\end{figure*}

The chemical potential of the barrier, if it were disconnected from the chain (the remaining $M-1$ sites of the ring), is the 
energy cost ${E_{\text{barrier}}(k+1) - E_{\text{barrier}}(k)}$ for adding a particle, 
given that the occupation is~$k$. 
Similarly we define the chemical potential of the chain, 
given that its occupation floor is~$q$. The explicit expressions are 
\beq
\mu_{\text{barrier}}(k) \ &=& \ Uk + W  \\
\mu_{\text{chain}}(q) \ &=&  \ Uq - 2J_q 
\eeq  
The first expression takes into account the on-site energy~$W$ of the barrier site, 
while the second expression takes into account the offset~$-2J_q$ due to the formation
of a conducting band. The occupation floor $q$ leads to a renormalized hopping amplitude $J_q$, see Section~V, and therefore to a wider band \hide{[as (maybe) will be seen in the spectrum figs once ill make a better fig.]}. It follows that the resonance condition 
for the transfer of the $k+1$ particle from the chain to the barrier is 
\beq
U \, k + W\ \ - \Delta = \ \ U\, q_{k+1} - 2J_q  
\eeq
The additional detuning parameter $\Delta$ is required because 
the on-site energy of the barrier is affected by the coupling 
to the quasi-continuum states of the chain. This shift  
could be ignored if the barrier site were weakly coupled.
From the resonance condition we deduce the equation for the $k$th line 
in the regime diagram, namely,  
\beq
\label{e5}
U \ = \ \frac{W+(2J_q-\Delta)}{q_{k+1}-k}  
\eeq
with $k=0,1,...,q_c$.
In \Fig{fg1} these line are indicated. In fact we neglect there 
the $2J_q-\Delta$ shift, and still the agreement is very good.
In Section~V we further discuss the determination of $\Delta$.

An analogous effect is also present in simply connected geometries, where persistent currents are not possible. The combination of strong interactions and on-site potentials then leads to resonant tunneling. This has been studied in a linear BH chain with a barrier \cite{PhysRevB.91.054515}, as well in two site \cite{lee2008many,folling2007direct,PhysRevA.78.031601}, three site \cite{1367-2630-12-6-065020} and superlattice models \cite{PhysRevLett.101.090404,super_lat1,super_lat2}.

It is illuminating to inspect the energy spectrum at the vicinity of the resonances.
In \Fig{fg2} we show the low-energy spectrum for a system of $M=N=6$ particles. This is complementary to panel (d) of \Fig{fg1}.
For $w<u/N$, corresponding to the weak-barrier condition   $W < U $ (left panel) we observe a gap in the spectrum as in the standard Mott insulating state, yielding  an exponentially suppressed persistent current.  For $w=u/N$ (middle panel) the gap closes, and instead we have an energy band of $6$ states, leading to large current(note the different axis limits on middle panel). For $w > u/N$ (right panel) there is no gap. However, the current is again strongly suppressed -- in this case the  large barrier effectively  disconnects the ring and reduces its size to an effective lattice  of $M-1$ sites. The particle removed from the barrier is delocalized all through the lattice, forming  a $5$-state band.

In the upper panels of \Fig{fg2a} we plot the persistent current $\mathcal{I}$ as a function of $u$ at fixed $w$; correspondingly in the lower panels  we plot the current  $\mathcal{I}$ as a function of $w$  at fixed $u$. At difference from \Fig{fg1} we do not plot the maximal persistent amplitude, rather its value for a fixed value of the flux $\Phi=0.9 \pi$. The reason is that in the same figure we also display  the average barrier occupation $\langle n_1 \rangle$, which is well defined only once the flux value is fixed.
The figure shows  a clear correlation between the barrier occupancy,  which displays a staircase behaviour, and the value of the current: the latter  displays a sharp increase when the occupancy of the barrier changes by one unit.
The expected location of the peaks, predicted in \Eq{e5}, is also plotted as red vertical dashed lines. The agreement is satisfactory, still,  the peak is slightly  shifted. A qualitative explanation for the shift will be provided in the next section.

\section{An effective single particle picture}

Having shown that the change by one unit of the barrier occupation is the origin of the resonant behaviour of the current, we further proceed by providing an effective single particle model near a transition line. In this case we effectively have $k$ and $(M-1)q$  "frozen" bosons at the barrier and at the chain respectively,  and $p_k$ bosons "free" to move along the ring. The ring feels now an effective barrier of $\tilde{W}=W-U(q-k)$. Note that $\tilde{W}$ can take both positive and negative values, and in the latter case the barrier acts effectively as a potential well.
In addition, the matrix elements of the kinetic part of the Hamiltonian, i.e. ${a}_{j{+}1}^{\dag} {a}_{j} $ lead to a factor of $ \sqrt{n_j(n_{j+1}+1)} $  when operating on a Fock state ${|   \{ n_j \}  \rangle}$ . This results in an effective hopping amplitude  $J_q=(q+1)J$ between the sites of the ring, except those connecting the barrier, where we have a smaller hopping amplitude of $J_w=\sqrt{(q+1)(k+1)}J$.
Considering for example the case of a single "free" particle ($p_k=1$), the problem can be described by the single particle Hamiltonian:
\begin{equation}
\label{e6} 
\mathcal{H}_{\text{sp}} \ = \tilde{W} | 1 \rangle  \langle 1 | -   \sum_{j=1}^{M} \tilde{J}_j \eexp{i(\Phi/M)}  | j+1 \rangle \langle j |
+ \text{h.c.} ,
\end{equation}
where $\tilde{J}_j=J_q$ for $j\neq 1$  and  $\tilde{J}_1=\tilde{J}_M=J_w$.
Using this model we calculate the single-particle current. The result is shown in the right panels of \Fig{fg2a} as dashed green lines. We see a perfect agreement between the current calculated using \Eq{e6} and the one obtained by exact diagonalization of the full Hamiltonian.

We can finally comment on the $2J_q-\Delta$ shift of the resonances in \Eq{e5}. In the absence of this correction, the expected location of the resonance, using \Eq{e5} is shown by red vertical lines in \Fig{fg2a}. In the effective single particle model this condition is simply given by $\tilde{W}=0$, which means that the ring in \Eq{e6} has no barrier nor a well. 
In \Fig{fg2a} we clearly see that the peak current is slightly shifted from the red lines.
This result is surprising, since one might expect that any potential barrier or a well (deviation from $\tilde{W}=0$) will only reduce the current. This claim would obviously be true for a translationally invariant ring. But the ring in $\mathcal{H}_{\text{sp}}$ is not translationally invariant due to the two weak links $J_w$. In this case a small potential well ($\tilde{W}<0$) actually increases the current.

The correction $2J_q-\Delta$ can be calculated as follows. The first part, $2J_q$, is an offset due to the formation of a conductance band in the chain.
In the absence of the barrier site, the Hamiltonian of the remaining $M-1$ sites ($j=2,..,M$) is diagonalized by:
\beq
 | n \rangle \ = \ \sqrt{\frac{2}{M}} \sum_{j=2}^M \sin \left( \frac{\pi n (j-1) }{ M } \right) | j \rangle 
\eeq
with the energies $\epsilon_n = -2J_q \cos(\pi n /M) $. 
The second part, $\Delta$, is a self-energy correction. The barrier site is connected to sites $j=2$ and $j=M$, so that the coupling $ V_n \equiv \langle j=1 | \mathcal{H}_{\text{sp}} | n \rangle  $ to the $n$-th mode of the chain is given by:
\beq
  V_n \ &=& \ -J_w \sqrt{\frac{2}{M}} \left[  \sin \left(\frac{\pi n }{M}\right) + \eexp{i\Phi} \sin \left(\frac{\pi n (M-1)}{M}\right) \right] \nonumber \\ 
   \ &=& \    -J_w \sqrt{\frac{2}{M}}  \sin \left( \frac{\pi n }{M}   \right)  \left[ 1 - (-1)^n  \eexp{i\Phi} \right]
\eeq
where for simplicity we have gauged the flux to the bond connecting the barrier and the $M$-th site.
We can now use second-order perturbation theory to calculate $\Delta$:
\beq   
\Delta \ = \ \sum_{n=1}^{M-1} \frac{|V_n|^2}{\epsilon_n-\tilde{W}} 
\eeq
when the energy of the effective barrier is approximately at the bottom of the band, namely $\tilde{W} \approx -2J_q$ we have:
\beq   \label{delta}
\Delta  \ &=& \ \frac{J_w^2}{J_q M} \sum_{n=1}^{M-1}  \frac{\sin^2 \left( \frac{\pi n }{M}   \right)\left| 1 - (-1)^n  \eexp{i\Phi} \right|^2 }{1-\cos \left(\frac{n\pi	}{M}\right)} \nonumber \\ 
\ &=& \ \frac{2 J_w^2}{J_q M} \sum_{n=1}^{M-1} \left( 1+ \cos \left(\frac{n\pi	}{M} \right) \right)  \left[ 1 - (-1)^n  \cos \Phi \right]  \nonumber \\ 
 \ &=& \  \frac{2J_w^2 (M-1+\cos \Phi)   }{J_q M}
\eeq
The expected location of the resonance, including the $2 J_q - \Delta$ correction is plotted in \Fig{fg2a} by vertical cyan lines. In the above treatment we have ignored the correction to the energy levels of the chain due to the barrier. This is justified for a large chain, therefore we expect this approximation to improve for larger $M$ values. In the large-$M$ limit, from \Eq{delta} we can see that the shift becomes independent on $\Phi$. In addition, in the absence of a weak link, i.e. $J_w=J_q$, and large $M$, we have $\Delta \rightarrow 2J_q$ so that the shift vanishes as expected.

\section{The effect of temperature}

\begin{figure}
\begin{center}

\includegraphics[width=0.95\hsize]{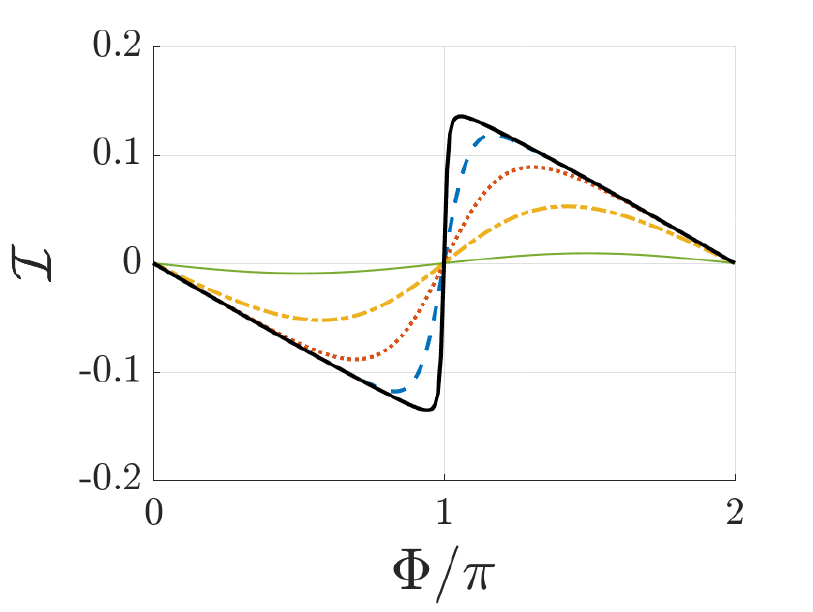}

\caption{\label{fg6}  
Finite-temperature persistent currents (in units of $2J$) 
for a ring of $M=6$ and $N=6$ as a function of the dimensionless flux $\Phi$. 
The dimensionless interaction and barrier strength are $u=3000$ and $w=495$. 
This corresponds to the center of the resonance that has been displayed 
in the upper right panel of \Fig{fg2a}.
The black line is the ground state current, while the colored lines correspond to different temperature values of: $k_B T/2J=0.04$ (dashed blue), $k_B T/2J=0.1$ (dotted red), $k_B T/2J=0.2$ (dashed-dot yellow), $k_B T/2J=0.5$ (thin solid green). 
} 
\end{center}
\end{figure}

\begin{figure}
\begin{center}
\includegraphics{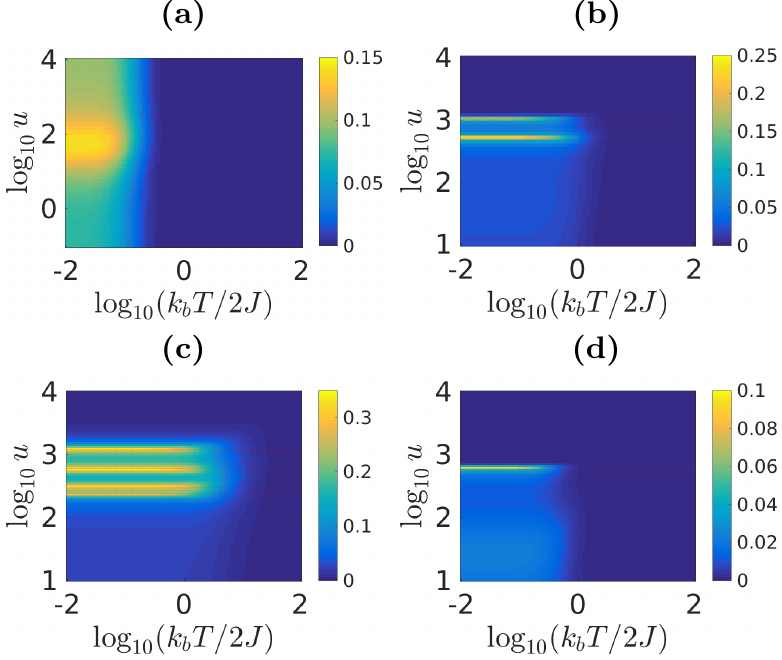} 
\caption{\label{fg5} Thermal averaged persistent current amplitude as a function of $u$ and $k_bT$. Panels (a)-(d) are for rings with $(M,N)$ as follows: $(10,4)$;$(5,10)$;$(3,12)$;$(6,6)$. In panel (a) the barrier strength is $w=10$ and in panels (b-d) $w=100$. The value plotted is the maximal canonical average current, in units of $2NJ/M$, picked from all $\Phi \in [0,\pi] $ with step size of $0.05 \pi$. The red dotted lines are given by \Eq{e5}.
} 
\end{center}
\end{figure}

After having studied in detail the zero-temperature regime-diagrams, we proceed to analyze the effects of thermal fluctuations on the system.
At finite temperature, the average current in the canonical ensemble is given by 
\begin{equation}
\bar{\mathcal{I}}  
\ =  \ \frac{\sum_i \mathcal{I}_i \eexp{-\beta E_i}}{\sum_i\eexp{-\beta E_i}}
\end{equation}
where $\beta$ is the inverse temperature. 

In \Fig{fg6} we display the finite-temperature persistent currents as a function of flux $\Phi$. We see that the zero temperature current is the largest for any value of $\Phi$, while thermal fluctuations reduce the persistent current amplitude, till it vanishes for $k_BT \simeq J $.

In \Fig{fg5} we plot the maximum value of the  current $\bar{\mathcal{I}}(\Phi)$  as a function of the temperature  and the reduced interaction $u$ for a fixed barrier height $w$ and different values of lattice size and filling. We see that, although  the temperature decreases the amplitude of persistent currents, the resonant effect remains visible at finite temperature  as long as persistent currents are non vanishing.

\section{Weak-link setup}

We consider separately a related setup that incorporates a ``weak link" instead of a barrier. 
In the lattice  tight-binding model a localized barrier is described by a local shift of the one-body potential~$W$ on one site, 
while  a weak link is obtained  by taking  a  smaller value for the tunnel amplitude~$J$ among two neighbouring sites. 
The latter choice  corresponds to larger mass in the continuum-limit, and not to a different potential.
We also note that \cite{Altman} disorder in $J$ leads to a ``Mott glass" phase, while disorder in $W$ leads to a ``Bose glass".   
The Hamiltonian for the weak-link case  is given by 
\begin{equation}
\mathcal{H} = \sum_{j=1}^{M} \left[
\frac{U}{2} {a}_{j}^{\dag} {a}_{j}^{\dag} {a}_{j} {a}_{j} 
- J_j \left(\eexp{i(\Phi/M)} {a}_{j{+}1}^{\dag} {a}_{j} + \text{h.c.} \right)
\right] \ \ \ \ \ 
\end{equation}
In our case there is no disorder but a single weak link, 
meaning that $J_1=J_w$ while $J_{i}=J$ otherwise.

\begin{figure}
\begin{center}

\includegraphics{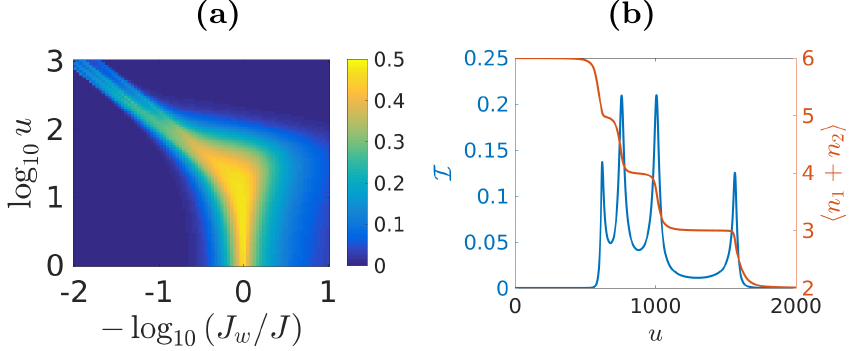}

\caption{ \label{fg7} 
Weak-link setup for a ring of $M=6$ and ${N=6}$. 
Note that we treat the strong link regime ($J_w/J>1$) on equal footing.
{\bf (a)} The persistent current amplitude $\alpha$  as a function of $u$ and $J_w/J$ (note the reversed x axis). The maximal current $I(\Phi)$ is picked from all $\Phi \in [0,\pi] $ with a step size of $0.05 \pi$. 
{\bf (b)} The ground state current (in units of $2J$), and the combined occupation of the two strongly connected sites, as a function of $u$, with fixed $J_w/J=80$ and $\Phi=0.9 \pi$.
}
\end{center}
\end{figure}

The naive thinking is to assume an analogy between the small~$J_w$ 
in this setup, and the large barrier~$w$ in the setup of \Fig{fg1}. 
For this reason we use a reversed sign in the horizontal axis of \Fig{fg7}a,
where we plot the persistent current amplitude as a function of $u$ and $J_w$. 
We see that qualitatively the results for ${N=M=6}$ look like those of \Fig{fg1}a, 
rather than like those of \Fig{fg1}d. 
This is because the weaker link does not lead 
to an expulsion of particles, hence no resonances arise.

We treat on equal footing the ``strong link" regime where $J_w/J>1$.
When we look on the left side of \Fig{fg7}a we observe 
a resonance structure similar to the case of having a barrier.
The explanation of the latter effect is as follows: we can regard the strong bond 
as a two-level system with energies $\pm J_w$. Ignoring the anti-bonding orbital, 
it is formally like having a ring with $\tilde{M}=M{-}1$ sites, and 
negative barrier ${W=-J_w}$. Thus we encounter resonances as discussed in previous sections.
The number of particles occupying the two strongly connected sites $n_1 + n_2$ has a minimal value of $2q_c + \min \{ p,2 \}$, obtained for large $U$ ($q_c$ and $p$ are defined as in \Eq{e9}). The term $ \min \{ p,2 \}$ is due to the bonding energy, which makes it energetically favourable to occupy also upto two of the excess particles. The maximal value of $n_1 + n_2$ is $N$, obtained for small $U$. Hence we have $N-2q_c - \min \{ p,2 \}$ resonances. An example is given in \Fig{fg7}b.

\section{Conclusions}

In conclusion, we have studied the persistent currents of a one-dimensional lattice ring, a theme which is related to the design of quantum simulators. Various strongly correlated phases can be addressed by tuning the barrier strength. The results of exact diagonalization have been provided for the persistent currents at various values of filling, as a function of the interaction, and of the barrier strength.  
At integer filling, the strong suppression  of persistent current amplitude generically signals the onset of the Mott insulator phase. We observe that when the barrier energy is of the order of interaction strength, a resonant behaviour occurs for the persistent currents, associated with the change of occupation of the site hosting the barrier. An effective single-particle model well accounts for the main observed features.
Some subtle differences between having a ``barrier", as opposed to  ``weak link" and ``strong link", have been highlighted.   
The effect of finite temperature is to decrease the persistent current amplitude, 
but the observed resonances remain robust as long as the persistent current does not vanish.
In outlook, it would be interesting to address larger system sizes, 
in order to study the interplay of thermal and quantum phase and density fluctuations on the system.


{\bf Acknowledgments.-- }
We acknowledge financial support from the ANR SuperRing (ANR-15-CE30-0012-02)




%


\onecolumngrid

\ \\ \ \\ \ \\ 

\Cn{
\parbox{0.6\linewidth}{
{\large
{\bf Physical Review A highlight.-- }  Persistent currents are studied for a one-dimensional bosonic gas in a ring lattice that has a SQUID-like geometry. While generically such currents are expected to be strongly suppressed in the Mott insulator phase, they show a pronounced resonant behavior when the barrier strength ($w$) becomes of the order of the inter-particle interaction energy ($u$).}}
}

\ \\

\Cn{\includegraphics[width=9cm]{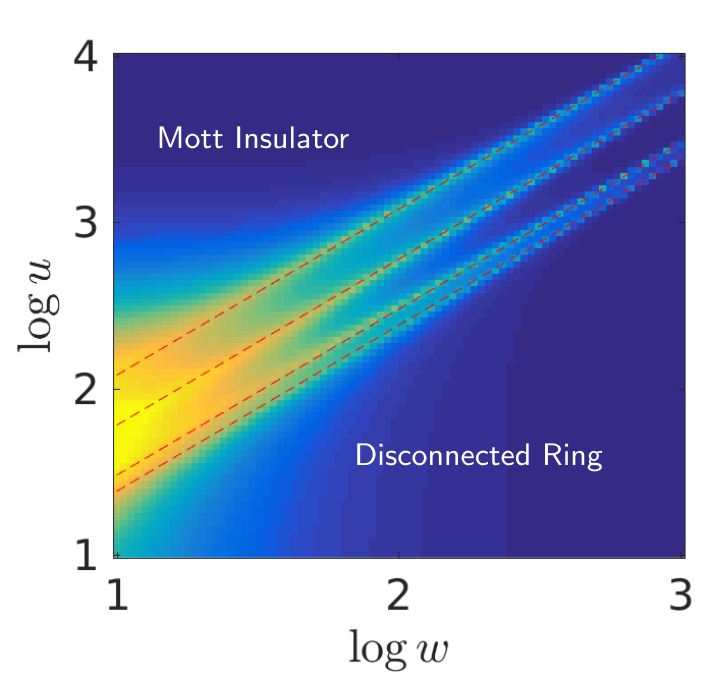}}

\clearpage
\end{document}